\PassOptionsToPackage{unicode}{hyperref}
\PassOptionsToPackage{hyphens}{url}
\documentclass[twocolumn,aps,prb,amsmath,amssymb,color,longbibliography,superscriptaddress]{revtex4-2}
\usepackage{amsmath,amssymb}
\usepackage{lmodern}
\usepackage{mathtools}
\usepackage{iftex}
\ifPDFTeX
  \usepackage[T1]{fontenc}
  \usepackage[utf8]{inputenc}
  \usepackage{textcomp} 
\else 
  \usepackage{unicode-math}
  \defaultfontfeatures{Scale=MatchLowercase}
  \defaultfontfeatures[\rmfamily]{Ligatures=TeX,Scale=1}
\fi
\IfFileExists{upquote.sty}{\usepackage{upquote}}{}
\IfFileExists{microtype.sty}{
  \usepackage[]{microtype}
  \UseMicrotypeSet[protrusion]{basicmath} 
}{}
\makeatletter
\@ifundefined{KOMAClassName}{
  \IfFileExists{parskip.sty}{%
    \usepackage{parskip}
  }{
    \setlength{\parindent}{0pt}
    \setlength{\parskip}{6pt plus 2pt minus 1pt}}
}{
  \KOMAoptions{parskip=half}}
\makeatother
\usepackage{xcolor}
\usepackage{graphicx}
\makeatletter
\def\maxwidth{\ifdim\Gin@nat@width>\linewidth\linewidth\else\Gin@nat@width\fi}
\def\maxheight{\ifdim\Gin@nat@height>\textheight\textheight\else\Gin@nat@height\fi}
\makeatother
\setkeys{Gin}{width=\maxwidth,height=\maxheight,keepaspectratio}
\makeatletter
\def\fps@figure{htbp}
\makeatother
\ifLuaTeX
  \usepackage{selnolig}  
\fi
\IfFileExists{bookmark.sty}{\usepackage{bookmark}}{\usepackage{hyperref}}
\IfFileExists{xurl.sty}{\usepackage{xurl}}{} 
\hypersetup{
  hidelinks,
  pdfcreator={LaTeX via pandoc}}

\date{\today}

\begin{document}
\begin{titlepage}
\title{Revisiting the Formulation of Charged Defect in Solids}
\author{Hanzhi Shang}
\author{Zeyu Jiang}
\author{Yiyang Sun} 

\author{D West}
\author{SB Zhang}
\affiliation{Department of Physics, Applied Physics and Astronomy, Rensselaer Polytechnic Institute, Troy, NY, 12180, USA}

\begin{abstract}

Defect physics is at the heart of microelectronics. By keeping track of the reference energy in total energy calculations, we explicitly show that the ``potential alignment'' correction vanishes, and the classic Markov-Payne correction yields accurate results. From linear response theory, we further formulate an accurate expression for the quadrupole correction. Application to numerous defects including anisotropic material yields accurate formation energies in small supercells and the historically slow convergence of the 2+ diamond vacancy is shown to be a result of slow varying gap levels of the defect leading to a size dependent dielectric constant.

\end{abstract}
\maketitle
\draft
\vspace{2mm}
\end{titlepage}
Defects play an important role in semiconductor microelectronics \cite{ref1}. With recent development of quantum materials, their roles are only further enhanced \cite{ref2}. In the past, first-principles calculations have played a pivotal role in identifying physical properties of the defects such as their formation energy and level positions inside the band gap. However, such calculations are almost exclusively performed within the periodic supercell approximation in which a compensating jellium charge background must be used for charged defects to eliminate the diverging Coulomb repulsion between supercells. As a result, the approach runs into problems of fictitious interaction between the defect and its periodic images, as well as with the compensating jellium.

The defect formation energy is defined as \cite{ref3}
\begin{equation}
\Delta H_f = \Delta E_{tot} + \sum_{i}^{}{n_i\mu_i} + q( E_{VBM} + E_F )
\end{equation}
where $\Delta E_{tot} = E_{tot}(D^q) - E_{tot}(bulk)$ is the total energy difference of the supercell containing a defect in charge state $q$ and the defect-free bulk supercell, $\mu_i$ are the chemical potentials of the atomic reservoirs (to account for the fact that defect formation may not conserve numbers of particles in the supercell), $E_F$ is the Fermi level (which serves as the electron reservoir) and is measured from the top of the valence band ($E_{VBM}$). Notably, $\Delta E_{tot}$  depends on the supercell size $L$; only at the dilute limit (i.e., zero defect density), is Eq. (1) uniquely valued. For a finite supercell, all energy levels including $E_{VBM}$ can be significantly affected by the defect. Usually, bulk VBM is used to measure $E_F$ with the understanding that its position relative to the defect cell is shifted by an amount
$\Delta\bar{V} = \bar{V}_B -\bar{V}_D$, where ${\bar{V}}_{B,D}$ represents the energy of an electron at the average electrostatic potential for bulk and defect supercell, respectively. In theory, if the cell size is ``large enough'' one can find a region in the defect supercell most close to bulk to calculate $\Delta\bar{V}$. The corresponding correction to the reservoir energy {[}i.e., the last term in Eq. (1){]} (potential alignment correction) would be $q\Delta \bar{V}$. A significant effort in correcting the supercell calculations can be traced to this aspect of the defect energy \cite{ref4}.


As both $\Delta E_{tot}$ and $q\Delta \bar V$ depend on the size (and shape) of the supercell used in the calculation, this inevitably affects the predicted $\Delta H_f$ and the defect transition level positions, hindering our ability to understand defects. To correct for the finite-size errors in $\Delta E_{tot}$, a multipole expansion in terms of a monopole Madelung term (order $\frac{1}{L})$ and a quadrupole term (order $\frac{1}{L^3}$) is usually applied, but details may vary especially in terms of the quadrupole. Alternatively, potential based methods require $\sl{ad\ hoc}$ modeling of the defect charge density to account for difficult-to-converge defects, such as the (2+) vacancy in diamond \cite{ref5}. Recently, it was even questioned whether some 30+ years of first-principles defect calculations is on the correct path. Citing an earlier work \cite{ref6} on vacancies in Si, it was suggested that the convergence of $\Delta H_f$ with respect to supercell size is very slow if not impossible. Instead, one should place the compensating charge in the conduction band edge for positive charged defects instead of as a jellium, although such a view is yet to gain traction \cite{ref7,ref8}.

In this paper, we present an ab-initio, simple to implement, finite-cell size charged defect correction based on the linear dielectric response of bulk, without requiring fitting or modeling of the defect charge density. To do this, first we reformulate the defect formation energy wherein we track the effects of a common reference energy \cite{ref9}. We show that the potential alignment correction should not be included if one uses the standard Markov-Payne (MP) type correction. Next, we propose an unambiguously defined quadrupole correction determined from the charge difference between defect and bulk supercells. Calculations of point defects in diamond, Si, MgO, and $\rm{Ga_2O_3}$ allow us classify them into 3 categories: (i) those that maintain the bulk electron bonding network, such as a substitutional defect, for which a Madelung correction is sufficient, (ii) those, such as an interstitial or vacancy, for which a quadrupole correction is necessary, and (iii) those, such as a vacancy in diamond, which are extremely slow to converge ($\frac{1}{L}$) due to size-dependent dielectric screening. Even for the diamond vacancy, our correction in the 64-atom supercell leads to a result that is converged to within 90 meV.

Without loss of generality, consider here an elemental solid. Within density functional theory, the total energy of a supercell  was established as \cite{ref10},
\begin{multline}
E_{tot} = \Sigma_i E_i - \frac{\Omega}{2}\Sigma_{\textbf{G} \neq 0}V_{Coul}( \textbf{G} )\eta( \textbf{G} ) \\
- \frac{\Omega}{4}\Sigma_{\textbf{G}}\mu_{xc}( \textbf{G} )\eta( \textbf{G} ) + \gamma_{Ewald} + \alpha_1 Z
\end{multline}
where $E_i$ are the eigenvalues, $\Omega$ is the volume, $V_{Coul}$ is the Coulomb potential, $\mu_{xc}$ is the exchange-correlation potential, $\eta( \textbf{G} )$ is the Fourier-transformed electron density, $\gamma_{Ewald}$ is the Ewald energy, $Z$ is the total ion charge, and $\alpha_1 = \Omega^{-1}\int_{}^{}{( U_{ps}(r) + \frac{2Z}{rN} )d^{3}r}$, where $U_{ps}$ is the pseudopotential and $N$ is the number of atoms in the supercell. Use of Eq. (2) for charged systems requires caution. The last two terms in Eq. (2) originate from,
\begin{equation}
E_0=\lim_{\textbf{G} \rightarrow 0}\Omega (\frac{1}{2} V_{Coul}(\textbf G) + U_{ps}(\textbf G)) \eta (\textbf G) +\frac{1}{2}\sideset{}{'}\sum_{\nu} \frac{2 Z^2}{|\textbf R_\nu|} 
\end{equation}
where the first term is electron-electron interaction, the second is electrons in the pseudopotential of the ions, and the final sum is ion-ion interaction energy. While previously this final sum is taken assuming the average potential of the ions is zero, such that $\frac{1}{2}\sideset{}{'}\sum_{\nu} \frac{2 Z^2}{|\textbf R_\nu|}  = \gamma_{Ewald} + \frac{1}{2}\lim_{\textbf{G}\rightarrow 0} \frac{8\pi Z_I^2}{ \Omega G^{2}}$, here we formally expand the electron and ion charge densities in order to keep track of the reference potentials. For small $\textbf G$, the electron density $\eta(\textbf G) \approx \frac{Z_e}{\Omega} + \beta_eG^2$ and similarly the ion density $\eta_I(\textbf G) \approx \frac{Z_I}{\Omega}+\beta_IG^2$. From the Poisson's equation, 
\begin{equation}
\begin{aligned}
\bar{V}_e &= \lim_{G\rightarrow0} 8\pi \frac{\eta(\textbf G)}{G^2}=8\pi\beta_e+\lim_{G\rightarrow0}\frac{8\pi}{G^2}\frac{Z_e}{\Omega}
\\
\bar{V}_I &=- \lim_{G\rightarrow0} 8\pi \frac{\eta_I(\textbf G)}{G^2}=-8\pi\beta_I-\lim_{G\rightarrow0}\frac{8\pi}{G^2}\frac{Z_I}{\Omega}
\end{aligned}
\end{equation}
are the energy of an electron in the average potential of electron(ion) charge densities. Note that both diverge due to the monopole term, with the total average $\bar V= \bar V_I + \bar V_e = 8\pi (\beta_e-\beta_I) + \lim_{G\rightarrow0}\frac{8\pi}{G^2}\frac{Z_e-Z_I}{\Omega}$ being finite when the system is charge neutral.
As detailed in the SM, Eq. (3) can be rewritten as
\begin{equation}
E_0=\gamma_{Ewald} + \alpha_1 Z_e + (Z_e - Z_I)\bar V - \lim_{G\rightarrow 0}\frac{4\pi}{\Omega G^2}(Z_e-Z_I)^2
\end{equation}
While this indeed reduces to the well known result of $\alpha_1 Z + \gamma_{Ewald}$ for charge neutral systems for which $Z_e=Z_I=Z$. Note here that for non-neutral systems both of the last two terms diverge independently (and when summed together). As such, the practice of using Eq. (2) to determine the energy of a "charged defect" by neglecting these divergent terms always corresponds to charge neutral calculation which implicitly includes a compensating jellium such that $Z_e \rightarrow Z_e + q =Z_I=Z$, where $q$ is the charge state of the defect and $Z_e$ is the explicit number of valence electrons. As the system is charge neutral, any shift in the  average electrostatic potential $\Delta \bar V$ has no contribution to the  total energy $(Z_e + q - Z_I)\Delta \bar V = 0$. 

The advantage of Eq. (5) is that it makes apparent the role of the average potential in the formation energies of Eq. (1). As the average electrostatic potential of periodic calculations are typically set to zero, electrons which are not explicitly considered in the calculation can be considered to have an energy of the reference energy, i.e., its average electron potential energy. For this reason, we write
\begin{equation}
E_{tot}(D^q)_{\bar V_D}=E_{tot}(D^q) + (Z_e^D + q -Z_I^D)\bar V_D
\end{equation}
where  $E_{tot}(D^q)$ is the total energy calculated with the average potential set to zero, the superscript $D$ indicates $Z_e$ and $Z_I$ are for the defect supercell, and $\bar V_D$ in the final term is taken relative to the vacuum level. 
Similarly, for bulk we have
\begin{equation}
E_{tot}(bulk)_{\bar V_B}=E_{tot}(bulk) + (Z_e^B - Z_I^B)\bar V_B
\end{equation}
where $\bar V_B$  is the average potential for the bulk supercell with respect to the same common vacuum level as $\bar V_D$ and the superscript $B$ indicates $Z_e$ and $Z_I$ are for bulk.
Explicitly using a common references for the two calculations, Eq. (1) can then be written as
\begin{equation}
\Delta H_f = E_{tot}(D^q)_{\bar V_D} - E_{tot}(bulk)_{\bar V_B}+ \sum_i n_i \mu_i,
\end{equation}
where the chemical potential for jellium electrons is implicitly $\bar V_D$. Using Eqs. (6)-(8), we have
\begin{multline}
\Delta H_f =E_{tot}(D^q)-E_{tot}(bulk) + (Z_e^D + q - Z_I^D)\bar V_D \\
 - (Z_e^B-Z_I^B)\bar V_B + \sum_i n_i \mu_i
\end{multline}

Over the years, various approximations \cite{ref4,ref5,ref6,ref7,ref8,ref11,ref12,ref13,ref14,ref15,ref16,ref17,ref18} have been made to remove unphysical interactions in $\Delta E_{tot}$ associated with finite cell size leading to interactions between defects as well as with jellium. In the classic Markov-Payne multipole expansion approach for a cubic supercell of length $L$ \cite{ref11}, the unphysical interactions \cite{ref19} (maintaining Rydberg atomic units in which $e^{2} = 2$)are given by
\begin{equation}
E_{int}^{q}(L) = - \frac{q\alpha}{L}\widetilde{q} + \frac{4\pi q}{3L^3}{\widetilde{Q}}_{loc} + O( L^{- 5} )
\end{equation}
where $\alpha$ is the Madelung constant, $\widetilde{q} = q/\varepsilon$, and ${\widetilde{Q}}_{loc} = \frac{1}{\varepsilon}\int_{\Omega}^{\ }{r^{2}\rho_{c}( \textbf{r} )d^{3}r}$ is the screened quadrupole moment due to localized defect charge $\rho_{c}( \textbf{r} )$. Here we note, however, that as the charge densities of the defect supercell, $\rho_D(\textbf r)$, and the bulk supercell, $\rho_B(\textbf r)$, differ, $\Delta \rho (\textbf r)= \rho_D(\textbf r)-\rho_B(\textbf r)$, there is an average potential associated with $\Delta \rho (\textbf r)$ of $\bar V_{\Delta \rho}= \bar V_D - \bar V_B$. Similar to Eqs. (6) and (7), the reference of the unphysical interaction energy can be explicitly included to give,  
\begin{equation}
(E_{int}^{q})_{\bar V_{\Delta \rho}}=E_{int}^{q} +((Z_e^D-Z_e^B )+q-(Z_I^D-Z_I^B ))\bar V_{\Delta \rho}  
\end{equation}
where $E_{int}^{q}$ is unphysical charged defect interaction calculated with the average potential set to zero. To correct for the errors introduced due to unphysical interactions in a finite cell, the corrected formation energy, $\Delta H_f^C$ is obtained by subtracting $(E_{int}^{q} )_{\bar V_{\Delta \rho}}$ from Eq. (9), to yield
\begin{multline}
\Delta H_f^C =
E_{tot}(D^q)-E_{tot}(bulk)-E_{int}^{q}\\
+(Z_e^D+q-Z_I^D ) \bar V_B + \sum_i n_i \mu_i ,
\end{multline}
where we have taken $Z_e^B$ to be $Z_I^B$. The inclusion of $-E_{int}^q$ in $\Delta H_f^C$ has not only removed the unphysical interaction between defects but has also explicitly resulted in a shift of the reference potential of jellium from $\bar V_D$ to $\bar V_B$. Hence, when putting the jellium electrons at the electron reservior in determination of the formation energy in the dilute limit, 
\begin{multline}
\Delta H_f^C =E_{tot}(D^q)-E_{tot}(bulk)-E_{int}^{q}\\
+ \sum_i n_i \mu_i+q( E_{VBM} + E_F),
\end{multline}
where $E_{tot}(D^q)$ and $E_{tot}(bulk)$ are both calculated in the usual manner, with their average potentials set to zero, and $E_{VBM}$ is relative to the bulk average potential.

Proceeding with applying the MP model, we note that it was developed rigorously for a localized defect charge density $\rho_{c}( \textbf{r} )$ in the absence of screening, which is inaccessible to first-principles calculation. Here instead, we use $\Delta \rho (\textbf{r})$ as defined earlier, which includes the dielectric response of the host material,
\begin{equation}
\Delta \rho( \textbf{r} )  = - \Delta \eta( \textbf{r} ) + \Delta q_{I} \delta( \textbf{r} ) - \frac{q}{\Omega}
\end{equation}
where $\Delta\eta( \textbf{r} ) = \eta_{D}( \textbf{r} ) - \eta_{B}( \textbf{r} )$ is the difference in the explicit electron density of defect and bulk supercells, $\Delta q_{I} = q_{D} - q_{B}$ is the difference in ion charge number for a point defect located at the origin, and $-q/\Omega$ is the jellium within the defect supercell. Following MP, before screening (bs) takes place $\Delta \rho ^{bs} ( \textbf{r} ) = \rho_c(\textbf{r}) - \frac{q}{\Omega}$,  and from linear response theory, 
\begin{equation}
\Delta \rho( \textbf{r} ) = \frac{1}{\varepsilon}(\rho_c(\textbf{r}) - \frac{q}{\Omega}). 
\end{equation}
Combining Eqs. (14) and (15) leads to the expression for the screened quadrupole 
\begin{equation}
\widetilde Q_{loc} = - \int_{\Omega}^{\ }{r^{2}\Delta\eta( \textbf{r} )d^{3}r} - \frac{q}{\Omega}\int_{\Omega}^{\ }r^2 (1-\frac{1}{\varepsilon}) d^{3}r .
\end{equation}

We apply the above approaches to defects in diamond, silicon, and cubic MgO. Our supercells are based on 8-atom cubic unit cells, e.g., a $4 \times 4 \times 4$ cell contains 512 atoms, which, for simplicity, are fixed at their respective ideal positions. Calculations were performed using density functional theory, as implemented in the VASP code \cite{ref20}. We used the Perdew-Burke-Ernzerhof (PBE) functional \cite{ref21} for exchange and correlation. The cutoff energies were 400 eV for diamond, MgO and $\text{F}_\text{i}^-$ in Si, and 300 eV for $\text{P}_\text{Si}^+$ and $\text{Al}_\text{Si}^-$ in Si. The lattice constants of diamond, silicon, and MgO used in calculations are 3.57 Å, 5.47 Å, and 4.20 Å. The self-consistent charge density was determined using a convergence criteria of $10^{-5}$ eV. As we must deal with variable cell size, the k-point mesh for Brillouin zone was tested and it was determined that $L(Å)N \geq 20$, where $N$ is the number of $k$ points in one direction, is sufficient when the $\Gamma$ point was avoided.
\begin{figure}[tbp]
	\includegraphics[width=1.00\columnwidth]{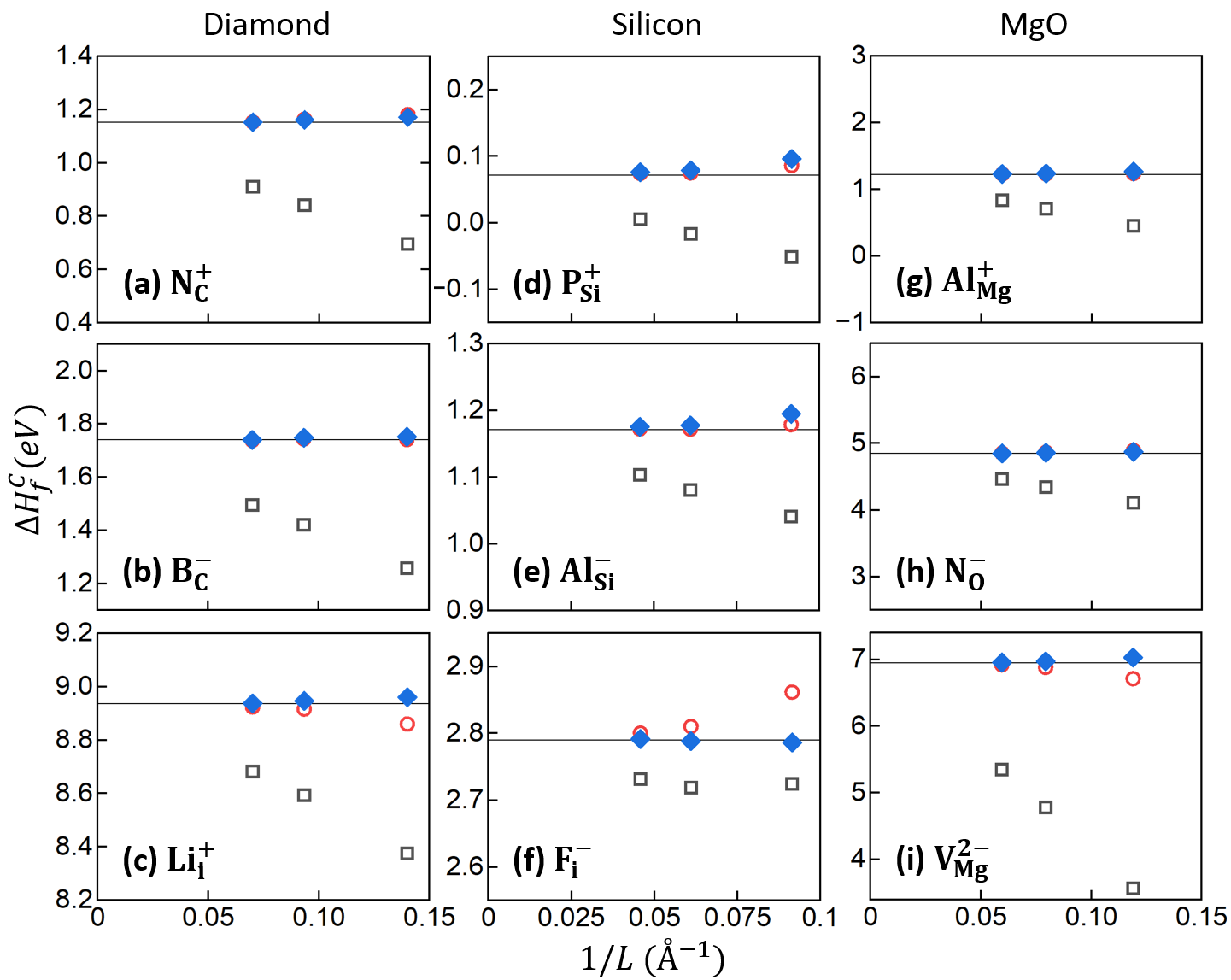}
	\label{fig:fig1}
	\caption{Corrected defect formation energy in diamond, silicon, and MgO with varying cell size. Open black squares are uncorrected formation energies. Red open circles are those with Madelung correction, while filled blue diamonds have both Madelung and quadrupole correction applied.}
\end{figure}

Figure 1 summarizes results for corrected formation energy $\Delta H_f^C$. The first two rows are substitutional defects which preserve the host electron bonding network. As such, the electron density is quite similar to that of bulk and the quadrupole associated with these defects is also small. Therefore, a Madelung correction is sufficient to remove the unphysical interactions, when compared with the extrapolated, i.e., converged, value even for the smallest 64-atom cell. The last row in Fig. 1 involves interstitial and vacancy for which the electron bonding network is noticeably altered. As a result, the quadrupole correction can be substantially larger, up to 0.3 eV for $\text{V}_\text{Mg}^{2-}$ in a  64-atom cell, bringing our results into good agreement with the converged values.
\begin{figure}[tbp]
	\includegraphics[width=1.00\columnwidth]{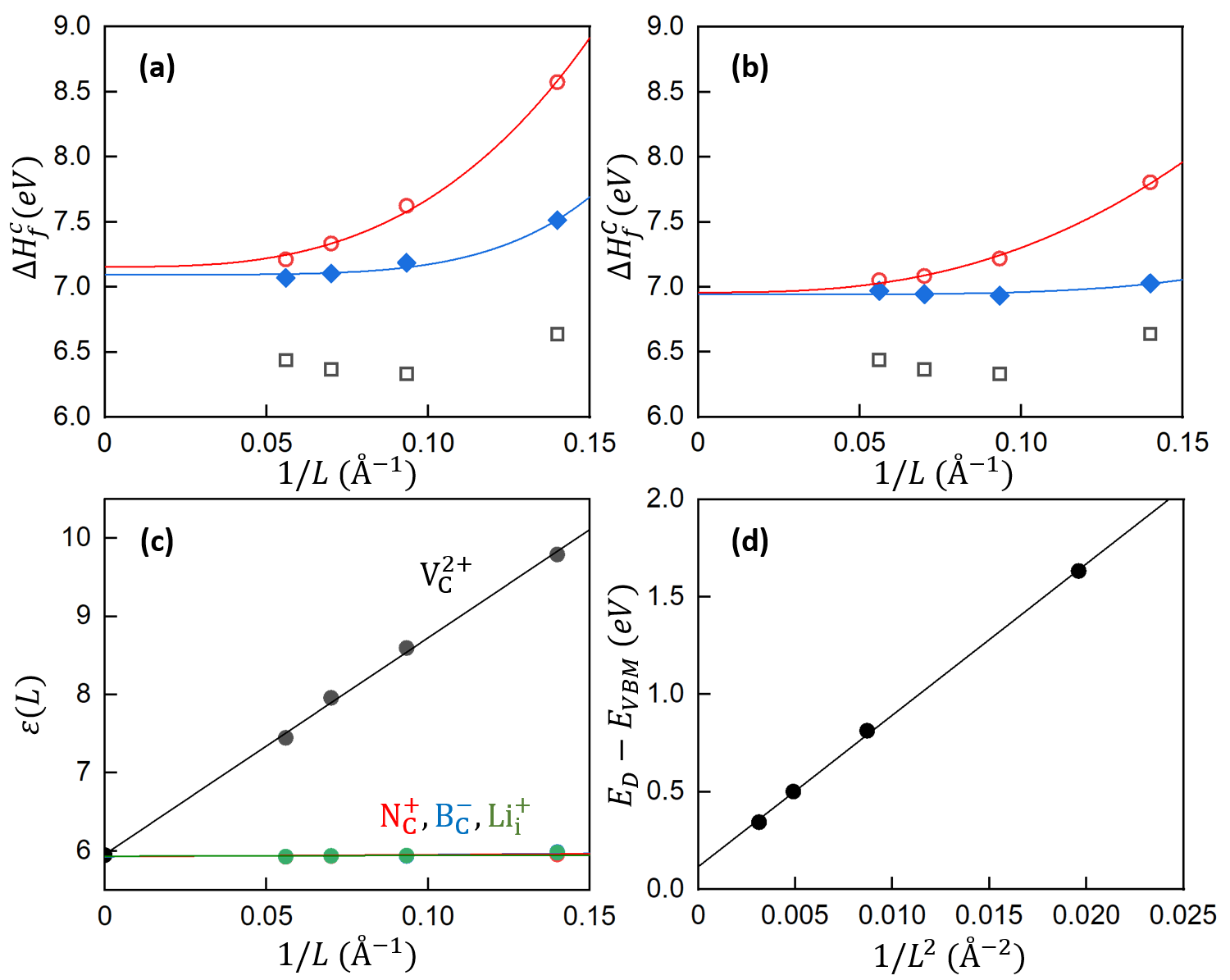}
	\label{fig:fig2}
	\caption{$\text{V}_\text{C}^{2+}$. (a) Formation energy using bulk $\varepsilon$, uncorrected (open black squares), with Madelung correction (open red circles), and with Madelung + quadrupole correction (solid blue diamonds), (b) formation energy with $\varepsilon(L)$ incorporated, (c) DFPT calculated $\varepsilon(L)$ for defect supercells of size $L$, and (d) empty defect $t_2$ state inside the band gap at $\Gamma$ point, as a function of $1⁄L^2$. Fit lines in (c) and (d) are linear and in (a) and (b) are detailed in the text.}
\end{figure}

Although Fig. 1 illustrates the success of our approach, the 2+ vacancy in diamond highlighted in Fig. 2 indicates the importance using the correct value of  $\varepsilon$. Here, we find that the quadrupole correction is exceptionally large (\textasciitilde10 times that of $\text{Li}_\text{i}^+$), yet when using $\varepsilon_{bulk}$ in Eq. (16) the correction is not sufficient to remove the remaining error, as $\Delta H_f^C$ for a 64-atom cell is still more than 0.4 eV larger than the converged value. Furthermore, fitting the expected error in Fig. 2(a) to $O(L^{- 3})$ and $O(L^{- 5})$ for Madelung and Madelung + quadrupole, respectively, we find that the results do not even approach the same limit. We stress that as $\widetilde q$ and $\widetilde Q$ which enter $E_{int}$ are associated with the screening in the defect supercell and as such $\varepsilon(L)$ is an explicit function of the supercell size $L$. In contrast to Fig. 2 (a), Fig. 2(b) shows that when using $\varepsilon(L)$, calculated using density functional perturbation theory (DFPT), values obtained from the 64-atom cell are within 0.09 eV of the large supercell limit.  This distinction in $\varepsilon$ is important for $\text{V}_\text{C}^{2+}$ as $\varepsilon(L)$ is substantially different from the bulk value of 5.9 even for supercells with more than 1,000 atoms, as revealed by DFPT calculation [see Fig. 2(c)]. This is in startling contrast to other defects, e.g., $\text{N}_\text{C}^+$, $\text{B}_\text{C}^-$, and $\text{Li}_\text{i}^+$, in Fig. 2(c). 

\textit{Why does $\varepsilon$ depend on cell size?} To understand the defect influence on $\varepsilon$(L), we consider electronic screening $\varepsilon(L)= \varepsilon_{bulk} + \varepsilon_{D}$, where the first term is the bulk contribution and the second term is the defect contribution. As detailed in the SM, the defect contribution can be written as \cite{ref22}
\begin{equation}
 \varepsilon_D = \frac{16\pi}{\Omega}\sum_{n,\textbf{k}}^{}{\sum_{i}^{}\frac{w_{\textbf{k}}| \langle \Psi_{D\textbf{k}}^{(i)}| \hat{q} \cdot \textbf{r} |\Psi_{n\textbf{k}} \rangle |^2}{E_{D\textbf{k}} - E_{n\textbf{k}}}} \propto \frac{1}{\Omega}
\end{equation}
where $i$ stands for the degeneracy of defect states, $D$, and the sum over $n$ is over occupied states. For $\text{V}_\text{C}^{2+}$, there are only four relevant defect states, one is the occupied $a_1$ singlet, whereas the other three are the empty $t_2$ triplet inside the band gap. As the contribution to $\varepsilon$ is sensitive to the energy separation $E_{D\textbf{k}} - E_{n\textbf{k}}$, here we consider only transitions from occupied bulk $p$ states to the empty $t_2$ states, whose energy is rather close to the VBM, see for example, $E_D - E_{VBM}$ in Fig. 2(d). For simplicity, we will further assume that $E_{D\textbf{k}} - E_{n\textbf{k}}$ may be approximated by transitions from an \emph{average} bulk $p$ state to the $t_2$ states, i.e., $\Delta E_{D,p} = E_D - {\bar{E}}_p$, then $\varepsilon_D \propto \frac{1}{\Omega\Delta E_{D,p}}$. If $\Delta E_{D,p}$ does not change with cell size, $\varepsilon_D$ will fall off with $\frac{1}{\Omega}$, which is exactly what one might expect from the Clausius--Mossotti relation for a gas of isolated atoms/molecules \cite{ref23,ref24,ref25}.

However, for $\text{V}_\text{C}^{2+}$, one cannot consider $\Delta E_{D,p}$ to be a constant. Figure 2(d) shows $E_D - E_{VBM}$ as a function of $\frac{1}{L^2}$, which reveals a surprisingly good linear scaling, and approaches a rather small value as $\frac{1}{L^2} \rightarrow 0$. This result reminds us of the work of Froyen and Harrison \cite{ref26} who showed that the strength of interatomic interactions in a cubic solid exhibit a $\frac{1}{L^2}$ dependence. This same functional dependence for defect level position can be derived (as shown in the SM), by considering the interaction of the $t_2$ states with the low lying $a_1$ state and all their images in different supercells. For the case where transitions from the band edge are the dominate contributions to $\varepsilon_D$, $\varepsilon_D \propto \frac{1}{\Omega\Delta E_{D,p}}$ $\propto 1/L$ and the dielectric constant of the supercell can be approximated as $\varepsilon(L) \approx \varepsilon_{bulk} + C/L$. Figure 2(c) reveals that this is indeed the case for the range of supercell sizes considered here, where the data is well represented by the least squares fit line of $\varepsilon(L) = \varepsilon_{bulk} + 27.7 \AA/L$.

\textit{Extension to nonisotropic systems}. While we have focused on the cubic cell for simplicity, it is important to note that the method presented here can be generalized to non-cubic and anisotropic supercells. The leading Madelung correction term for such systems has already been derived by  Rurali and Cartoixà and applied to defect systems \cite{ref27,ref28}. From the Green's function of a point charge in a anisotropic dielectric \cite{ref29}, in the anisotropic case the screened quadrupole in Eq. (16) has the same functional form, except that $\varepsilon$ is replaced by $\varepsilon_{eff}$ with the expression,
\begin{equation}
 \frac{1}{\varepsilon_{eff}} = \frac{1}{4\pi \sqrt{det(\bar\varepsilon)}}\int_0^{\pi}d\theta sin\theta \int_0^{2\pi} \frac{d\phi }{\sqrt{\hat r^T \bar\varepsilon^{-1} \hat r}}
\end{equation}
where $\bar \varepsilon$ is the dielectric tensor associated with the supercell. Detailed derivation of Eq. (18) and its application to Ga\textsubscript2O\textsubscript3 can be found in the SM.

In summary, we have presented an linear response theory based correction for finite cell size effects associated with calculating the formation energies of charged defects in semiconductors which is simple to implement and requires neither fitting nor an assumed functional form for the defect charge distribution. Application to a number of charged defects in diamond, Si, MgO, and Ga\textsubscript2O\textsubscript3, show the formation energies are well converged for even the smallest 64-atom supercells considered. This includes the historically difficult to converge $\text{V}_\text{C}^{2+}$, where we show that the origin of its slow convergence can be attributed to the strong dependence of $\varepsilon$ on the size of the defect supercell.

This work was supported by the U.S. DOE Grant No. DE-SC0002623. The supercomputer time sponsored by the National Energy Research Scientific Center (NERSC) under DOE Contract No. DE-AC02-05CH11231 and that by Advanced Cyberinfrastructure Coordination Ecosystem: Services \& Support (ACCESS), under NSF grant number ACI-1548562. SBZ also acknowledges the computational resources from Purdue Anvil and TAMUFASTER supercomputers made available by ACCESS through allocation DMR180114, as well as the Center for Computational Innovations (CCI) at RPI.

{*}Current address: Shanghai Institute of Ceramics, CAS, 215 Chengbei Rd, Shanghai, China.


\begin{thebibliography}{30}
\bibitem{ref19}Based on the definition of $E_{12}$ in Eq. (14), it appears the sign for the quadrupole term in Eq. (15) should be positive.
\end{thebibliography}
\end{document}